\documentclass[10pt,a4paper,onecolumn]{article}
\usepackage{marginnote}
\usepackage{graphicx}
\usepackage{xcolor}
\usepackage{authblk,etoolbox}
\usepackage{titlesec}
\usepackage{calc}
\usepackage{tikz}
\usepackage{hyperref}
\hypersetup{colorlinks,breaklinks=true,
            urlcolor=[rgb]{0.0, 0.5, 1.0},
            linkcolor=[rgb]{0.0, 0.5, 1.0}}
\usepackage{caption}
\usepackage{tcolorbox}
\usepackage{amssymb,amsmath}
\usepackage{ifxetex,ifluatex}
\usepackage{seqsplit}
\usepackage{xstring}

\usepackage{float}
\let\origfigure\figure
\let\endorigfigure\endfigure
\renewenvironment{figure}[1][2] {
    \expandafter\origfigure\expandafter[H]
} {
    \endorigfigure
}

\usepackage{fixltx2e} 
\usepackage[
  backend=biber,
]{biblatex}
\bibliography{bibliography.bib}

\let\textttOrig=\texttt
\def\texttt#1{\expandafter\textttOrig{\seqsplit{#1}}}
\renewcommand{\seqinsert}{\ifmmode
  \allowbreak
  \else\penalty6000\hspace{0pt plus 0.02em}\fi}

\makeatletter
\let\href@Orig=\href
\def\href@Urllike#1#2{\href@Orig{#1}{\begingroup
    \def\Url@String{#2}\Url@FormatString
    \endgroup}}
\def\href@Notdoi#1#2{\def\tempa{#1}\def\tempb{#2}%
  \ifx\tempa\tempb\relax\href@Urllike{#1}{#2}\else
  \href@Orig{#1}{#2}\fi}
\def\href#1#2{%
  \IfBeginWith{#1}{https://doi.org}%
  {\href@Urllike{#1}{#2}}{\href@Notdoi{#1}{#2}}}
\makeatother

\newlength{\cslhangindent}
\newlength{\csllabelwidth}
\newenvironment{CSLReferences}[3] 
 {
  \setlength{\parindent}{0pt}
  \ifodd #1 \everypar{\setlength{\hangindent}{\cslhangindent}}\ignorespaces\fi
  \ifnum #2 > 0
  \setlength{\parskip}{#2\baselineskip}
  \fi
 }%
 {}
\usepackage{calc}

\usepackage[top=3.5cm, bottom=3cm, right=1.5cm, left=1.0cm,
            headheight=2.2cm, reversemp, includemp, marginparwidth=4.5cm]{geometry}

\titleformat{\section}
  {\normalfont\sffamily\Large\bfseries}
  {}{0pt}{}
\titleformat{\subsection}
  {\normalfont\sffamily\large\bfseries}
  {}{0pt}{}
\titleformat{\subsubsection}
  {\normalfont\sffamily\bfseries}
  {}{0pt}{}
\titleformat*{\paragraph}
  {\sffamily\normalsize}

\usepackage{fancyhdr}
\makeatletter
\let\ps@plain\ps@fancy
\fancyheadoffset[L]{4.5cm}
\fancyfootoffset[L]{4.5cm}

\definecolor{linky}{rgb}{0.0, 0.5, 1.0}

\newtcolorbox{repobox}
   {colback=red, colframe=red!75!black,
     boxrule=0.5pt, arc=2pt, left=6pt, right=6pt, top=3pt, bottom=3pt}

\newcommand{\ExternalLink}{%
   \tikz[x=1.2ex, y=1.2ex, baseline=-0.05ex]{%
       \begin{scope}[x=1ex, y=1ex]
           \clip (-0.1,-0.1)
               --++ (-0, 1.2)
               --++ (0.6, 0)
               --++ (0, -0.6)
               --++ (0.6, 0)
               --++ (0, -1);
           \path[draw,
               line width = 0.5,
               rounded corners=0.5]
               (0,0) rectangle (1,1);
       \end{scope}
       \path[draw, line width = 0.5] (0.5, 0.5)
           -- (1, 1);
       \path[draw, line width = 0.5] (0.6, 1)
           -- (1, 1) -- (1, 0.6);
       }
   }

\patchcmd{\@maketitle}{center}{flushleft}{}{}
\patchcmd{\@maketitle}{center}{flushleft}{}{}
\patchcmd{\@maketitle}{\LARGE}{\LARGE\sffamily}{}{}
\def\maketitle{{%
  
  \AB@maketitle}}
\makeatletter
\renewcommand\AB@affilsepx{ \protect\Affilfont}
\renewcommand\AB@affilnote[1]{{\bfseries #1}\hspace{3pt}}
\renewcommand{\affil}[2][]%
   {\newaffiltrue\let\AB@blk@and\AB@pand
      \if\relax#1\relax\def\AB@note{\AB@thenote}\else\def\AB@note{#1}%
        \setcounter{Maxaffil}{0}\fi
        \begingroup
        \let\href=\href@Orig
        \let\texttt=\textttOrig
        \let\protect\@unexpandable@protect
        \def\thanks{\protect\thanks}\def\footnote{\protect\footnote}%
        \@temptokena=\expandafter{\AB@authors}%
        {\def\\{\protect\\\protect\Affilfont}\xdef\AB@temp{#2}}%
         \xdef\AB@authors{\the\@temptokena\AB@las\AB@au@str
         \protect\\[\affilsep]\protect\Affilfont\AB@temp}%
         \gdef\AB@las{}\gdef\AB@au@str{}%
        {\def\\{, \ignorespaces}\xdef\AB@temp{#2}}%
        \@temptokena=\expandafter{\AB@affillist}%
        \xdef\AB@affillist{\the\@temptokena \AB@affilsep
          \AB@affilnote{\AB@note}\protect\Affilfont\AB@temp}%
      \endgroup
       \let\AB@affilsep\AB@affilsepx
}
\makeatother

\renewcommand\Affilfont{\sffamily\small\mdseries}
\ifnum 0\ifxetex 1\fi\ifluatex 1\fi=0 
  \usepackage[T1]{fontenc}
  \usepackage[utf8]{inputenc}

\else 
  \ifxetex
    \usepackage{mathspec}
    \usepackage{fontspec}

  \else
    \usepackage{fontspec}
  \fi
  \defaultfontfeatures{Ligatures=TeX,Scale=MatchLowercase}

\fi
\IfFileExists{upquote.sty}{\usepackage{upquote}}{}
\IfFileExists{microtype.sty}{%
\usepackage{microtype}
\UseMicrotypeSet[protrusion]{basicmath} 
}{}

\usepackage{hyperref}
\hypersetup{unicode=true,
            pdftitle={SelfEEG: A Python library for Self-Supervised Learning in Electroencephalography},
            pdfborder={0 0 0},
            breaklinks=true}
\let\addcontentslineOrig=\addcontentsline
\def\addcontentsline#1#2#3{\bgroup
  \let\texttt=\textttOrig\addcontentslineOrig{#1}{#2}{#3}\egroup}
\let\markbothOrig\markboth
\def\markboth#1#2{\bgroup
  \let\texttt=\textttOrig\markbothOrig{#1}{#2}\egroup}
\let\markrightOrig\markright
\def\markright#1{\bgroup
  \let\texttt=\textttOrig\markrightOrig{#1}\egroup}

\IfFileExists{parskip.sty}{%
\usepackage{parskip}
}{
\setlength{\parindent}{0pt}
\setlength{\parskip}{6pt plus 2pt minus 1pt}
}
\providecommand{\tightlist}{%
  \setlength{\itemsep}{0pt}\setlength{\parskip}{0pt}}
\ifx\paragraph\undefined\else
\let\oldparagraph\paragraph
\renewcommand{\paragraph}[1]{\oldparagraph{#1}\mbox{}}
\fi
\ifx\subparagraph\undefined\else
\let\oldsubparagraph\subparagraph
\renewcommand{\subparagraph}[1]{\oldsubparagraph{#1}\mbox{}}
\fi

\title{SelfEEG: A Python library for Self-Supervised Learning in
Electroencephalography}

        \author[1, 2, 3]{Federico Del Pup}
          \author[3]{Andrea Zanola}
          \author[2, 3]{Louis Fabrice Tshimanga}
          \author[3]{Paolo Emilio Mazzon}
          \author[2, 3, 4]{Manfredo Atzori}
    
      \affil[1]{Department of Information Engineering, University of
Padova, Via Gradenigo 6/b, 35131 Padova, Italy}
      \affil[2]{Department of Neuroscience, University of Padua, Via
Belzoni 160, 35121 Padova, Italy}
      \affil[3]{Padova Neuroscience Center, University of Padova, Via
Orus 2/B, 35129 Padova, Italy}
      \affil[4]{Information Systems Institute, University of Applied
Sciences Western Switzerland (HES-SO Valais), 2800 Sierre, Switzerland}
  \date{\vspace{-7ex}}

\begin{document}
\maketitle

\marginpar{

  \begin{flushleft}
  \sffamily\small

  {\bfseries DOI:} \href{https://doi.org/DOI unavailable}{\color{linky}{DOI unavailable}}

  \vspace{2mm}

  {\bfseries Software}
  \begin{itemize}
    \setlength\itemsep{0em}
    \item \href{N/A}{\color{linky}{Review}} \ExternalLink
    \item \href{https://github.com/MedMaxLab/selfEEG}{\color{linky}{Repository}} \ExternalLink
    \item \href{DOI unavailable}{\color{linky}{Archive}} \ExternalLink
  \end{itemize}

  \vspace{2mm}

  \par\noindent\hrulefill\par

  \vspace{2mm}

  {\bfseries Editor:} \href{https://example.com}{Pending
Editor} \ExternalLink \\
  \vspace{1mm}
    {\bfseries Reviewers:}
  \begin{itemize}
  \setlength\itemsep{0em}
    \item \href{https://github.com/Pending Reviewers}{@Pending
Reviewers}
    \end{itemize}
    \vspace{2mm}

  {\bfseries Submitted:} N/A\\
  {\bfseries Published:} N/A

  \vspace{2mm}
  {\bfseries License}\\
  Authors of papers retain copyright and release the work under a Creative Commons Attribution 4.0 International License (\href{http://creativecommons.org/licenses/by/4.0/}{\color{linky}{CC BY 4.0}}).

  \vspace{2mm}
  {\bfseries Preprint Policy}\\
  This work has been submitted to The Journal of Open Source Software for possible publication. More information about the journal Preprint Policy can be found at the following \href{http://joss.readthedocs.io/en/latest/}{\color{linky}{link}}.

  \end{flushleft}
}

\hypertarget{summary}{%
\section{Summary}\label{summary}}

SelfEEG is an open-source Python library developed to assist researchers
in conducting Self-Supervised Learning (SSL) experiments on
electroencephalography (EEG) data. Its primary objective is to offer a
user-friendly but highly customizable environment, enabling users to
efficiently design and execute self-supervised learning tasks on EEG
data.

SelfEEG covers all the stages of a typical SSL pipeline, ranging from
data import to model design and training. It includes modules
specifically designed to: split data at various granularity levels
(e.g., session-, subject-, or dataset-based splits); effectively manage
data stored with different configurations (e.g., file extensions, data
types) during mini-batch construction; provide a wide range of standard
deep learning models, data augmentations and SSL baseline methods
applied to EEG data.

Most of the functionalities offered by selfEEG can be executed both on
GPUs and CPUs, expanding its usability beyond the self-supervised
learning area. Additionally, these functionalities can be employed for
the analysis of other biomedical signals often coupled with EEGs, such
as electromyography or electrocardiography data.

These features make selfEEG a versatile deep learning tool for
biomedical applications and a useful resource in SSL, one of the
currently most active fields of Artificial Intelligence.

\hypertarget{statement-of-need}{%
\section{Statement of need}\label{statement-of-need}}

SelfEEG answers to the lack of Self-Supervised Learning (SSL) frameworks
for the analysis of EEG data.

In fact, despite the recent high number of publications (more than 20
journal papers in the last 4 years (\hyperlink{ref-DelPup2023}{Del Pup \& Atzori, 2023}), there are
currently no frameworks or common standards for developing EEG-based SSL
pipelines, contrary to other fields such as computer vision (see
\href{https://github.com/lightly-ai/lightly}{LightlySSL} or
\href{https://github.com/facebookresearch/vissl}{ViSSL}).

In the field of EEG data analysis, where it has been demonstrated that
SSL can improve models' accuracy and mitigate overfitting (\hyperlink{ref-eegrafiei}{Rafiei et
al., 2022}) (\hyperlink{ref-banville}{Banville et al., 2021}), the absence of a self-supervised
learning framework dedicated to EEG signals limits the development of
novel strategies, reproducibility of results, and the progress of the
field.

Thanks to selfEEG, researchers can instead easily build SSL pipelines,
speeding up experimental design and improving the reproducibility of
results. Reproducibility is a key factor in this area, as it enhances
the comparison of different strategies and supports the creation of
useful benchmarks.

SelfEEG was also developed considering the needs of deep learning
researchers, for whom this library has been primarily designed. For this
reason, selfEEG aims to preserve a high but easily manageable level of
customization.

\hypertarget{library-overview}{%
\section{Library Overview}\label{library-overview}}

SelfEEG is a comprehensive library for SSL applications to EEG data. It
is built on top of PyTorch (\hyperlink{ref-pytorch}{Paszke et al., 2019}) and it includes several
modules targeting all the steps required for developing EEG-based SSL
pipelines. In particular, selfEEG comprises the following modules:

\begin{itemize}
\tightlist
\item
  \textbf{dataloading}: a collection of functions and classes designed
  to support data splitting and the construction of efficient PyTorch
  dataloaders in the EEG context.
\item
  \textbf{augmentation}: a collection of EEG data augmentation functions
  and other classes designed to combine them in more complex patterns.
\item
  \textbf{models}: a collection of EEG deep learning models.
\item
  \textbf{losses}: a collection of self-supervised learning losses.
\item
  \textbf{ssl}: a collection of self-supervised learning algorithms
  applied to EEG analysis with highly customizable fit methods.
\item
  \textbf{utils}: a collection of utility functions and classes for
  various purposes, such as a PyTorch compatible EEG sampler and scaler.
\end{itemize}

\hypertarget{related-open-source-projects}{%
\section{Related open-source
projects}\label{related-open-source-projects}}

Despite several deep learning frameworks were developed for the analysis
of EEG data, a library focused on the construction of self-supervised
learning pipelines on EEG data is still not available to the best of our
knowledge, hindering the advancement of the scientific knowledge and the
progress in the field. A comprehensive review of open-source projects
related to neuroscientific data analysis is provided in (\hyperlink{ref-app13095472}{Tshimanga et
al., 2023}). Few examples are EEG-DL (\hyperlink{ref-eegdl}{Hou et al., Feb. 2020}) and
\href{https://github.com/torcheeg/torcheeg}{torchEEG}, which
characterized for their completeness and spread among the
neuroscientific community.

\hypertarget{future-development}{%
\section{Future development}\label{future-development}}

Considering how rapidly self-supervised learning is evolving, this
library is expected to be constantly updated by the authors and the
open-source community, especially by adding novel SSL algorithms, deep
learning models, and functionalities that can enhance the comparison
between different developed strategies. In particular, the authors plan
to continue working on SelfEEG during the next years via several ongoing
European and national projects.

\hypertarget{credit-authorship-statement}{%
\section{CRediT Authorship
Statement}\label{credit-authorship-statement}}

FDP: Conceptualization, Writing - Original Draft, Software -
Development, Software - Design, Software - Testing; AZ: Writing - Review
\& Editing, Software - design (dataloading and utils modules), Software
- Testing; LFT: Writing - Review \& Editing, Software - design
(dataloading and utils modules), Software - Testing; PEM: Technical
support, Writing - Review \& Editing, Software - Testing; MA: Funding
Acquisition, Project Administration, Supervision, Writing - Review \&
Editing.

\hypertarget{acknowledgements}{%
\section{Acknowledgements}\label{acknowledgements}}

This work was supported by the STARS@UNIPD funding program of the
University of Padova, Italy, through the project: MEDMAX. This project
has received funding from the European Union's Horizon Europe research
and innovation programme under grant agreement no 101137074 -
HEREDITARY. We would also like to thank the other members of the Padova
Neuroscience Center for their support during the project development.

\hypertarget{references}{%
\section*{References}\label{references}}
\addcontentsline{toc}{section}{References}

\hypertarget{refs}{}
\begin{CSLReferences}{1}{0}
\leavevmode\hypertarget{ref-banville}{}%
Banville, H., Chehab, O., Hyvärinen, A., Engemann, D.-A., \& Gramfort,
A. (2021). Uncovering the structure of clinical EEG signals with
self-supervised learning. \emph{Journal of Neural Engineering},
\emph{18}(4), 046020. \url{https://doi.org/10.1088/1741-2552/abca18}

\leavevmode\hypertarget{ref-DelPup2023}{}%
Del Pup, F., \& Atzori, M. (2023). Applications of self-supervised
learning to biomedical signals: A survey. \emph{IEEE Access}, 1--1.
\url{https://doi.org/10.1109/ACCESS.2023.3344531}

\leavevmode\hypertarget{ref-eegdl}{}%
Hou, Y., Zhou, L., Jia, S., \& Lun, X. (Feb. 2020). A novel approach of
decoding EEG four-class motor imagery tasks via scout {ESI} and {CNN}.
\emph{Journal of Neural Engineering}, \emph{17}(1), 016048.
\url{https://doi.org/10.1088/1741-2552/ab4af6}

\leavevmode\hypertarget{ref-pytorch}{}%
Paszke, A., Gross, S., Massa, F., Lerer, A., Bradbury, J., Chanan, G.,
Killeen, T., Lin, Z., Gimelshein, N., Antiga, L., \& others. (2019).
Pytorch: An imperative style, high-performance deep learning library.
\emph{Advances in Neural Information Processing Systems}, \emph{32}.
\url{https://doi.org/10.48550/arXiv.1912.01703}

\leavevmode\hypertarget{ref-eegrafiei}{}%
Rafiei, M. H., Gauthier, L. V., Adeli, H., \& Takabi, D. (2022).
Self-supervised learning for electroencephalography. \emph{IEEE
Transactions on Neural Networks and Learning Systems}.
\url{https://doi.org/10.1109/TNNLS.2022.3190448}

\leavevmode\hypertarget{ref-app13095472}{}%
Tshimanga, L. F., Del Pup, F., Corbetta, M., \& Atzori, M. (2023). An
overview of open source deep learning-based libraries for neuroscience.
\emph{Applied Sciences}, \emph{13}(9).
\url{https://doi.org/10.3390/app13095472}

\end{CSLReferences}

\end{document}